\documentclass[useAMS,usenatbib,usegraphicx]{mn2e}
\usepackage{times}
\newcommand{\aj}{AJ}                   % Astronomical Journal
\newcommand{\apj}{ApJ}                 % Astrophysical Journal
\newcommand{\apjl}{ApJ}                % Astrophysical Journal, Letters

\newcommand{\aap}{A\&A}                % Astronomy and Astrophysics
  
\newcommand{\aaps}{A\&AS}              % Astronomy and Astrophysics, Supplement
\newcommand{\mnras}{MNRAS}             % Monthly Notices of the RAS
\newcommand{\echa}{$\eta$~Cha} 

\newcommand{\kms}{km~s$^{-1}$}
\newcommand{\msun}{M$_{\small\sun}$}
\newcommand{\msunyr}{M$_{\small\sun}$~yr$^{-1}$}

\begin{document}
\title[Disk accretion in the halo of $\eta$ Cha]{Episodic disk accretion in the halo of the `old' Pre-Main Sequence cluster $\eta$ Chamaeleontis\thanks{This paper includes data gathered with the 6.5 meter Magellan Telescopes located at Las Campanas Observatory, Chile.}}
\author[S. Murphy et al.]{Simon~J.~Murphy$^1$\thanks{Email: murphysj@mso.anu.edu.au (SJM); w.lawson@adfa.edu.au (WAL); bessell@mso.anu.edu.au (MSB); daniel@mso.anu.edu.au (DDRB).}, Warrick~A.~Lawson$^2$, Michael~ S.~Bessell$^1$ and 
\newauthor
and Daniel D. R. Bayliss$^1$\\
$^1$ Research School of Astronomy and Astrophysics, The Australian National University, Cotter Road, 
Weston Creek ACT 2611, Australia \\
$^2$  School of PEMS, University of New South Wales, Australian Defence Force Academy, Canberra, ACT 2600, Australia}

\maketitle
\begin{abstract}
We present multi-epoch medium-resolution observations of two M4.5 candidate members in the halo of the $\sim$8~Myr $\eta$~Chamaeleontis open cluster. Over six months of observations both stars exhibited variations in their H$\alpha$ line profiles on timescales of days to months, with at least one episode of substantial activity attributable to accretion from a circumstellar disk. We derive an accretion rate $\sim$10$^{-8.7}$~\msunyr\ for this event, with a rate of $\sim$10$^{-10.6}$~\msunyr\ in quiescence. Episodic accretion like that observed here means existing surveys of accreting Weak-lined T-Tauri Stars in young clusters are likely incomplete and that gas dissipation timescales calculated from the fraction of accreting objects are underestimates.
\end{abstract}


\section{Introduction}

The open cluster $\eta$~Chamaeleontis is one of the closest (94.3~pc) and youngest ($\sim$8~Myr) stellar aggregates in the Solar neighbourhood. Nearby, isolated groups such as \echa\ are ideal laboratories for investigating the dynamical evolution of young star clusters, in particular the influence that dynamics have had on the evolution of protoplanetary disks. Young clusters show a steady decline in the number of stars having disks and signatures of accretion with age \citep{Mohanty05,Jayawardhana06}. By an age of $\sim$5~Myr, 90--95\% of all young cluster members have stopped accreting material at a significant rate, yet $\sim$20\% of objects retain enough dust in their disks to produce a mid-IR excess. The mechanism responsible for these two different  timescales is still uncertain \citep{Fedele10}. \echa\ has already been shown to have both an excess of stars with accretion and stars with detectable disks compared to clusters of similar age \citep{Haisch01,Sicilia-Aguilar09}. The cluster also appears to have an Initial Mass Function deficient in low-mass objects \citep{Lyo04}. \citet{Fedele10} is the latest study to point out that dynamical evolution has probably dispersed a large fraction of the low-mass members to radii beyond that currently surveyed \citep[see also][]{Moraux07,Murphy10}. Depending on the properties of the dispersed members, it is possible that the current membership may be biased towards IR-excess and strong H$\alpha$ emitting sources. By investigating the disk and accretion properties of any new dispersed members of \echa\, we can hope to gain a more unbiased view of the cluster as a whole, as well as addressing any influence dynamical interactions have had on disk evolution.

In \citet[][hereafter Paper~\textsc{i}]{Murphy10} we presented the results of our search for the putative halo of low-mass objects surrounding \echa. From photometry, proper motions and multi-epoch spectra we identified 4 probable and 3 possible members, including two M4.5 stars which exhibited large variations in their H$\alpha$ emission line strengths. These stars, 2MASS~08014860$-$8058052 (hereafter 2MASS~0801) and 2MASS~08202975$-$8003259 (hereafter 2MASS~0820) are the subject of this letter. 

\section{Multi-epoch spectroscopy}

As part of our continued investigation into \echa\ we obtained multi-epoch, medium resolution spectroscopy of 2MASS~0801 and 2MASS~0820 with the \textit{WiFeS} instrument at the ANU 2.3~m telescope (see Paper \textsc{i} for more details).  We obtained 9 observations of 2MASS~0801 over 2010~January--June and 13 epochs of 2MASS~0820. In addition to H$\alpha$ emission both stars recurrently showed He~\textsc{i} 5876~\AA, 6678~\AA\, and Na~\textsc{i}~D emission, often associated with accretion. Details of these observations are given in Table~\ref{table:obs}. 

\begin{table} 
\centering
\caption{Multi-epoch \textit{WiFeS} observations.}
\label{table:obs}
\begin{tabular}{l@{\hspace{10pt}}c@{\hspace{10pt}}c@{\hspace{10pt}}c@{\hspace{10pt}}c} 
\hline
UTC of& H$\alpha$ EW & $v_{10}$ width & RV & Other\\
observation & [\AA] & [\kms] & [\kms] &lines\\
 \hline
\multicolumn{4}{l}{\textbf{2MASS~0801$-$8058}}\\
2010 Jan 25 13:57 & $-$6 & 152 &13.7\\
2010 Jan 27 12:35 & $-$6 & 148 & 23.7\\
2010 Jan 28 12:11 & $-$6 & 153 & 17.8 & He \sc i\\
2010 Feb 19 11:45 & $-$20 & 346 & 21.3 & He \sc i\\
2010 Apr 28 09:04 & $-$6 & 176 & 15.7\\
2010 Apr 29 09:05 & $-$6 & 160 & 20.9\\
2010 Apr 30 10:15 & $-$27 & 324 & 23.0 & He \textsc{i}, Na \textsc{d}\\ 
2010 May 01 13:14 & $-$7 & 178 & 18.5\\
2010 Jun 03 09:05 & $-$7 & 159 & 20.0& \\
\multicolumn{4}{l}{\textbf{2MASS~0820$-$8003}}\\
2010 Jan 25 14:45 & $-$23 & 331 & 17.2& He \sc i\\
2010 Feb 19 10:28 & $-$40 & 425 & 16.5& He \sc i\\
2010 Feb 20 10:27 & $-$24 & 260 & 18.8& He \sc i\\
2010 Apr 27 12:05 & $-$27 & 238 & 14.1 & He \textsc{i}, Na \textsc{d}\\
2010 Apr 28 10:14 & $-$17 & 263 & 18.9\\
2010 Apr 28 15:23 & $-$15 & 262 & 18.7\\
2010 Apr 29 10:16 & $-$17 & 282 & 20.6\\
2010 Apr 29 15:05 & $-$16 & 275 & 21.4\\
2010 Apr 30 09:05 & $-$17 & 237 & 20.0\\
2010 Apr 30 11:30 & $-$15 & 224 & 17.9\\
2010 Apr 30 12:50 & $-$16 & 210 & 18.4\\
2010 May 01 12:02 & $-$17 & 262 & 17.6\\
2010 Jun 03 10:15 & $-$17 & 215 & 18.9\\
\multicolumn{4}{l}{\textbf{2MASS~0820$-$8003 (Magellan/\textit{MIKE})}}\\
2010 May 11 23:41 & $-$22 & 238 & 17.4 & He~\textsc{i}\\
\hline
\end{tabular} 
\end{table} 

\subsection{Accretion diagnostics}

\begin{figure}
   \centering
   \includegraphics[width=0.44\textwidth]{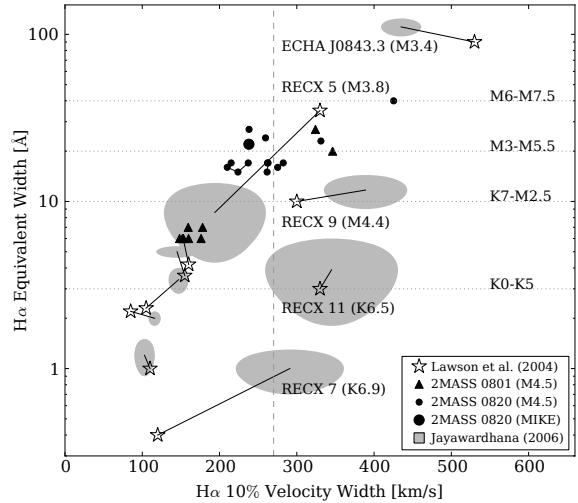} 
   \caption{H$\alpha$ Equivalent Width versus the H$\alpha$ velocity width at 10\% peak intensity ($v_{10}$) for 2MASS~0820 and 2MASS~0801. Horizontal lines denote the minimum EW for CTT stars in the indicated range of spectral types. The vertical line at 270 \kms\ separates accreting and non-accreting objects \citep{White03}. Shaded regions show the standard deviation of the multi-epoch measurements of \citet{Jayawardhana06}. Comparisons to the single epoch data of \citet{Lawson04} are also shown. Lines connect the two sets of measurements. For the 2MASS~0820 measurements lines connect observations taken on the same night.}
   \label{fig:ew_v10}
\end{figure}

To trace any accretion across our observations we analysed the time evolution of the H$\alpha$ equivalent width and velocity width of the H$\alpha$ line at 10\% intensity ($v_{10}$). 
% maybe take out this part, down to Figure...
Recently \citet{Nguyen09} have shown that due to the strong dependence of $v_{10}$ on the shape of the H$\alpha$ line profile, the Ca~\textsc{ii} 8662~\AA\ line flux is perhaps a more reliable quantitative diagnostic of accretion rate. The H$\alpha$ $v_{10}$ width has a well-established record as an accretion indicator and despite the caveats put forward by \citet{Nguyen09}, we use it here for comparison to previous studies and the fact that our wavelength coverage does not extend past 7100~\AA. 
Figure~\ref{fig:ew_v10} shows an EW-$v_{10}$ diagram for our observations of 2MASS 0801 and 0820, with \echa\ members observed by \citet{Jayawardhana06} and \citet{Lawson04} for comparison. Measurements of the EW and $v_{10}$ for each star are listed in Table~\ref{table:obs}. We estimate an uncertainty of $\pm$1~\AA\ for the EWs and $\pm$10 \kms\ in $v_{10}$, primarily due to uncertainties in defining the pseudo-continuum around the broad H$\alpha$ lines at this spectral resolution. Both stars show substantial variation in the EW-$v_{10}$ space, with the scatter more pronounced in 2MASS~0820.

 \citet{Lawson04} observed the cluster at high-resolution and found a similar EW variation of 2--3$\times$ for RECX 5, 9 and 11 when compared to the discovery measurements of \citet{Mamajek99}. \citet{Jayawardhana06} obtained 3--8 epochs of Magellan/\textit{MIKE} echelle data for 11 \echa\ members during 2004 December -- 2005 March. They report variations in EW and $v_{10}$ in some stars similar to those seen in 2MASS~0801 and 0820 with 35\% $v_{10}$ and 50\% EW variations in the maximal cases. In Figure~\ref{fig:ew_v10} we compare the measurements of these two groups. In general there is good agreement between the two sets of measurements, with RECX~9, RECX~11 and ECHA~J0843.3 all clearly accretors in both studies. RECX~7 is a known non-accreting SB2 spectroscopic binary \citep{Lyo03}, which explains the broad line width but low EW in the \citeauthor{Jayawardhana06} observations. \citet{Lawson04} classify RECX~5 as accreting with $v_{10}>300 \textrm{ \kms}$ from their 2002 observation. \citeauthor{Jayawardhana06} however failed to detect any accretion, with only one of their five observations showing a broadened H$\alpha$ line. The blue bump seen in the 2004 December spectrum could mean RECX~5 is sporadically accreting at low levels and \citeauthor{Lawson04} observed a strong outburst of 10$^{-10}$~\msunyr\ accretion in 2002, or the star is chromospherically active and underwent a period of strong activity during their observations. Chromospheric activity as a possible mechanism for the variations seen in our spectra is discussed in detail in \S\ref{sect:profiles}.

\begin{figure}
   \centering
   \includegraphics[width=0.43\textwidth]{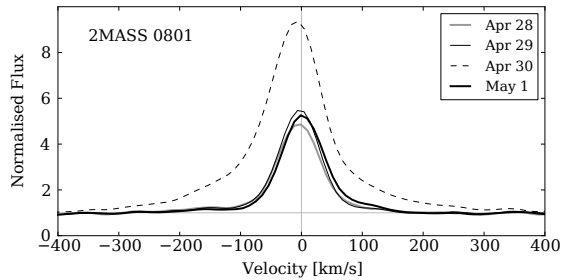} 
   \caption{Evolution of the 2MASS~0801 H$\alpha$ velocity profile during the 2010 Apr outburst. All profiles have been shifted to a radial velocity of zero. The Apr~30 profile has been veiling-corrected as described in the text.}
   \label{fig:april}
\end{figure}

Several sets of 2MASS~0820 spectra were taken during the same nights. They reveal a scatter of up to 25~\kms\ in $v_{10}$ and several Angstroms in EW, i.e. larger than the instrumental errors. While both stars generally lie on the non-accreting side of the 270~\kms\ criterion defined by \citet{White03}, they each make several excursions into the accreting region of the diagram, meeting both the EW and $v_{10}$ criteria for accreting Classical T-Tauri stars (CTTS). This is similar to the behaviour of RECX~5 during 2002--2005. The timescale of these excursions appears to be of the order hours to days. In the case of 2MASS~0801 both EW and $v_{10}$ increase dramatically on Apr~30, before returning to quiescent levels the next night. A similar level of activity is seen on 2010 Feb 19. Figure~\ref{fig:april} shows the evolution of the 2MASS~0801 H$\alpha$ profile during the 2010~April event. We obtained spectra before, during and after the event. The pre- and post- event line profiles are remarkably similar, with broad wings developing during the Apr~30 outburst, giving rise to a velocity width of $v_{10}>320$~\kms. The EW similarly quadruples to $-$27 \AA. The central velocity of the line profile also evolves with time (see Figure~\ref{fig:avg}). 2MASS~0820 appears to have 3 tiers of activity: a base level at $\textrm{EW}\approx-16$~\AA, a higher level at $-20>\textrm{EW}>-27$~\AA, then increasing to $-40$~\AA\ and $v_{10}=425$~\kms\ on 2010 Feb~19.

\begin{figure}
   \centering
   \includegraphics[width=0.43\textwidth]{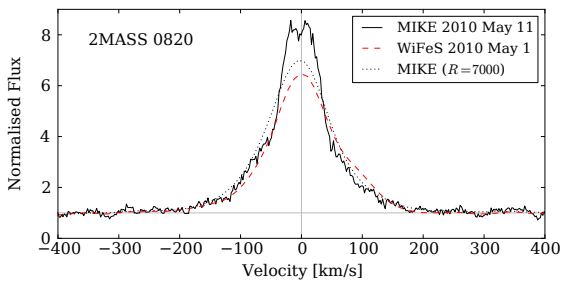} 
   \caption{\textit{MIKE} H$\alpha$ velocity profile of 2MASS~0820 compared to a contemporaneous \textit{WiFeS} observation.The dotted line shows the $R\approx25000$ \textit{MIKE} spectrum smoothed to the $R\approx7000$ resolution of \textit{WiFeS}.}
   \label{fig:mike}
\end{figure}

In addition to \textit{WiFeS} spectroscopy we obtained a Magellan/\textit{MIKE} spectrum of 2MASS~0820 on 2010 May 11. The H$\alpha$ velocity profile is shown in Figure~\ref{fig:mike} with a contemporaneous \textit{WiFeS} spectrum.  The EW and $v_{10}$ of the higher resolution spectrum agree well with the \textit{WiFeS} values (see Figure~\ref{fig:ew_v10}). The velocity profile shows some self-absorption in the line center as well as a slight excess of blueshifted emission at $v\approx-100$ \kms. To test how our lower resolution  \textit{WiFeS} spectra affect measurements of EW and $v_{10}$ we smoothed the \textit{MIKE} spectrum to $R\approx7000$ and re-measured $v_{10}$ and the EW. In both parameters the smoothed value was similar to that measured from the contemporaneous \textit{WiFeS} spectrum, i.e. smaller than the original value. We are thus likely underestimating the EW and $v_{10}$ from our \textit{WiFeS} observations.

\subsection{H$\alpha$ velocity profile variations}
\label{sect:profiles}

\begin{figure*}
   \centering
   \includegraphics[width=0.44\textwidth]{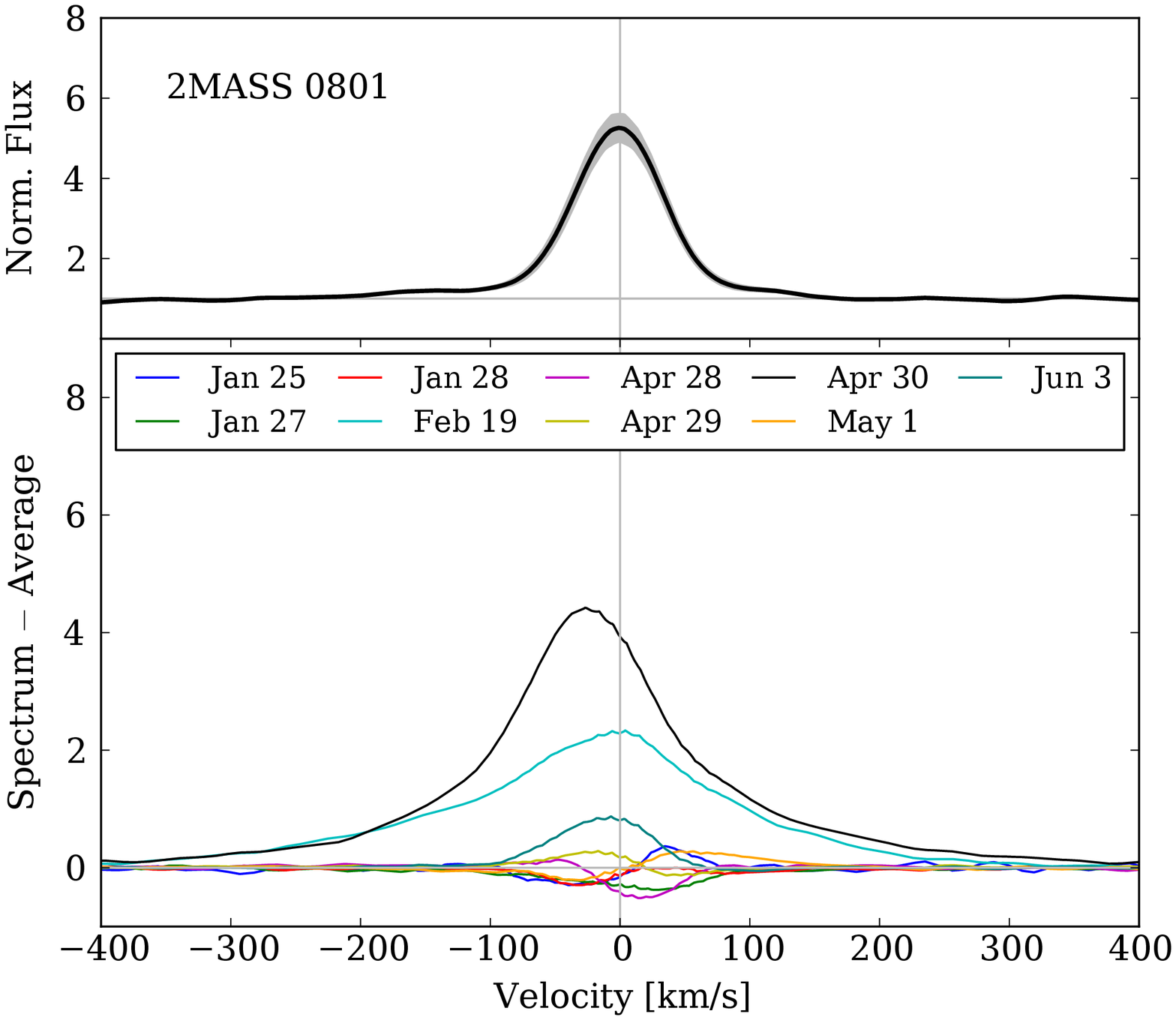} 
   \includegraphics[width=0.44\textwidth]{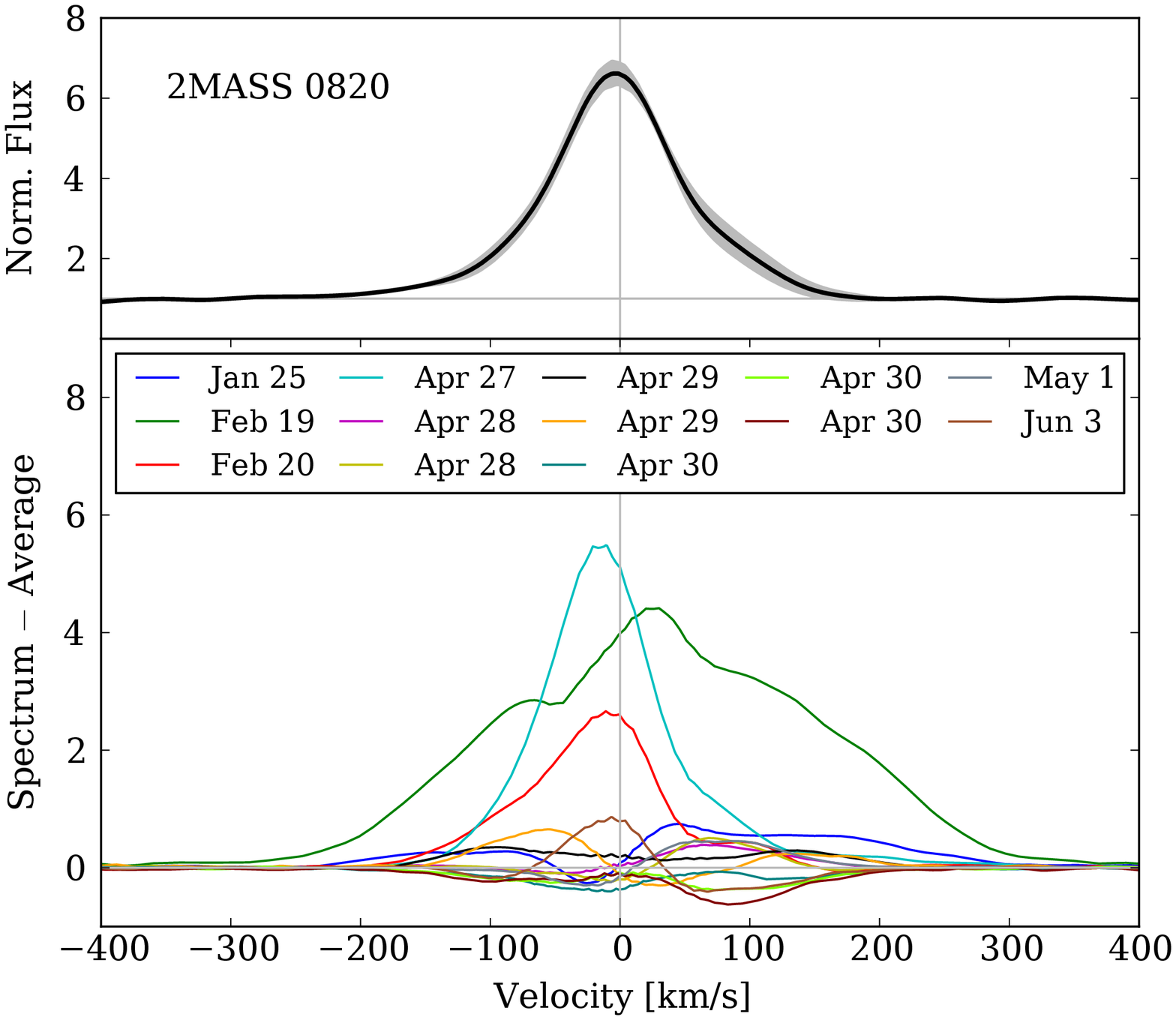} 
   \caption{\textit{WiFeS} H$\alpha$ velocity profiles for the 9 observations of 2MASS~0801 (left panel) and 13 observations of 2MASS~0820. In each plot the top panel shows the average quiescent spectrum spectrum and the standard deviation of quiescent spectra around the mean (shaded region). The bottom panel shows the variation around the mean for all epochs. Note the broad, multicomponent residual on Feb 19 for 2MASS~0820.}
   \label{fig:avg}
\end{figure*}

As well as the bulk quantities of H$\alpha$ EW and $v_{10}$ we also investigated the variation in the shape of the H$\alpha$ velocity profile over our observations. The top panels in Figure~\ref{fig:avg} show the average quiescent spectrum for each star, constructed using several (7--9) non-outburst spectra. The standard deviation of these spectra is also shown. 2MASS~0820 shows increased scatter around the mean profile at $v\approx+100$~\kms\ (and to a lesser extent at $-$100~\kms). This is due to a variable-strength red shifted component at this velocity present in some of the observations. The bottom panel shows the variation in the residual spectra -- the difference between the individual spectra and the mean spectrum. Immediately apparent are the broad residual profiles, tracing velocities up to $\pm$200--300~\kms. The Feb 19 epoch of 2MASS~0820 shows a residual velocity profile reaching to $\pm300$~\kms, with four distinct components visible and a large red asymmetry. Velocity shifts in the peaks of the residual spectra are present at up to several tens of \kms. 

Instead of invoking accretion, can chromospheric activity also explain the strong and varying H$\alpha$ profiles? The work of \citet{Montes98} has shown that in some WTTS the H$\alpha$ line profile cannot be fitted by a single gaussian and two components are necessary: a narrow gaussian of FWHM $<$100~\kms\ and a much broader component with FWHM 130--470~\kms, sometimes offset in wavelength from the narrow component. They attribute these line profiles to micro-flaring occuring in the chromospheres of the stars. Micro-flares are frequent, short duration events and have large-scale motions that could explain the broad wings observed in the lines and the residual spectra in Figure~\ref{fig:avg}. \citet{Stauffer97} observe a similar effect in several of their $\sim$30~Myr IC~2391/2602 and 100~Myr Pleiades targets. Both our stars show He~\textsc{i} 6678~\AA\ in emission at their peak H$\alpha$ levels. Strong He~\textsc{i} 6678~\AA\ emission is an accretion diagnostic as it is generally only present in low-levels ($\ll$1~\AA) in older chromospherically active stars \citep{Gizis02}. While we do detect strong (1.5~\AA) emission in the April~30 outburst spectrum of 2MASS~0801, at all other epochs where we detect the line it is weak ($\sim$0.5~\AA). Furthermore, \citet{Martin01} detect strong (1--4~\AA) He~\textsc{i} 6678~\AA\ emission during a flare of the old M9 dwarf LHS~2065. 

Given the weak He~\textsc{i} line strengths generally observed in our stars and the simple gaussian-like profiles of the residual spectra we do not have strong evidence for ongoing accretion. Chromospheric activity is a much more likely explanation for the observed line profiles and EW/$v_{10}$ variations. Only the February~19 spectrum of 2MASS~0820 shows a broad, asymmetric residual characteristic of accretion. Multiple components are present at velocities up to $\pm$300~\kms, presumably tracing the ballistic infall of material from the inner edge of the disk onto the stellar surface. 

\subsection{Accretion rates}

\citet{Natta04} have shown that a good correlation exists between the H$\alpha$ $v_{10}$ velocity width and the mass accretion rate across a large range of masses, 0.04--0.8~\msun. If we assume the H$\alpha$ emission is a result of accretion we can use the \citet{Natta04} relation to derive the mass accretion rate ($\dot{M}$) without the need to perform detailed model fits to the line profile. We derive a quiescent accretion rate for 2MASS~0801 of $\log\dot{M}=-11.3\pm0.3$~\msunyr, where the uncertainty reflects the variation in $v_{10}$ and uncertainties in the relation parameters. For 2MASS~0820 the quiescent scatter is larger and $\log\dot{M}=-10.6\pm0.5$~\msunyr. For the event of February 19 we calculate an accretion rate of $\log\dot{M}=-8.7\pm0.5$~\msunyr. This is similar to that derived for ECHA~J0843.3 from detailed H$\alpha$ profile modeling by \citet{Lawson04}. The \citet{Natta04} relation gives an accretion rate 1.3~dex larger for the corresponding $v_{10}$ velocity given by \citeauthor{Lawson04} According to \citet{Nguyen09} the large scatter in this average relation probably reflects object-to-object variation (possibly due to evolutionary effects) rather than the effects of variability on the usually single-epoch observations as previously thought \citep[e.g.][]{Scholz06}. As previously mentioned, the effect of our lower spectral resolution means we have likely underestimated $v_{10}$ so our derived accretion rates are probably even higher. 2MASS~0820 shows a 10--80$\times$ jump in accretion rate between quiescence and its active phase. Using the \citeauthor{Natta04} relation on the high-resolution RECX~5 velocities of \citet{Lawson04} and \citet{Jayawardhana06} results in a similar 20$\times$ change in accretion rate. Assuming the scatter in quiescent $v_{10}$ velocities is a result of the accretion rate varying with time, the variations are similar in magnitude to those seen in the 10~Myr CTTS TW Hya \citep{Eisner10} and lower-mass young brown dwarfs \citep{Scholz06,Stelzer07}. 

\subsection{Continuum veiling in 2MASS~0801$-$8058}

\begin{figure*}
   \centering
   \includegraphics[width=0.88\textwidth]{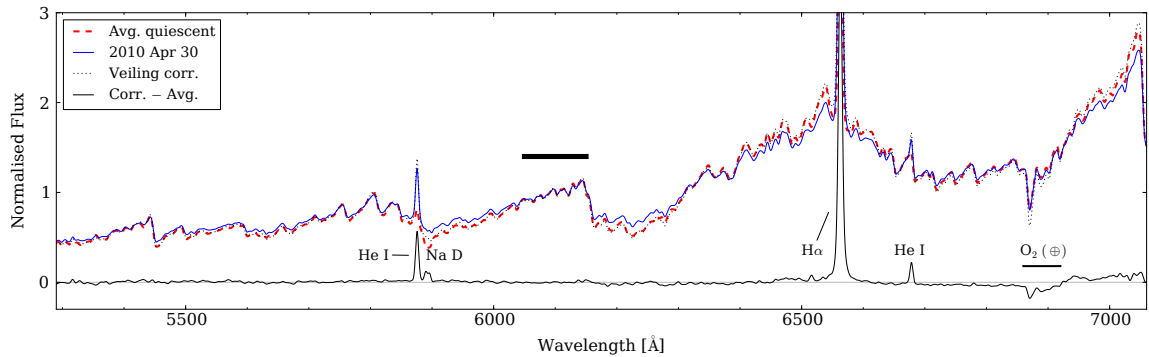} 
   \caption{The effects of continuum veiling on the 2010 Apr 30 spectrum of 2MASS~0801. All spectra smoothed by a 10~pixel Gaussian kernel and are normalised over the 6050--6150~\AA\ region (heavy line). The red dotted line shows the final veiling-corrected spectrum, as calculated by applying Equation\ \ref{eqn:veiling}.}
   \label{fig:veiling}
\end{figure*}

The April 30 spectrum of 2MASS~0801 exhibits a Li~\textsc{i}~6708~\AA\ equivalent width smaller than in quiescence. This is the effect of continuum veiling, where enhanced continuum emission fills in absorption lines and reduces the measured EW. Magnetospheric infall models predict the presence of veiling from accretion shocks, as matter from the accretion flow reaches the stellar surface \citep[e.g.][]{Muzerolle03}. However, this is probably not the case for 2MASS~0801.  Figure~\ref{fig:veiling} shows the full \textit{WiFeS} spectrum for April 30 compared to the average 2MASS~0801 quiescent spectrum. Veiling is readily apparent as an enhancement of the continuum in the 6200 \AA\ CaH band and a corresponding depression of the continuum around H$\alpha$ (due to the normalisation at 6100~\AA). Using the quiescent spectrum as a reference we find the veiling at 6100~\AA\ is $\sim$0.22. The observed flux $F$ was veiling-corrected using the following prescription:
\begin{equation}
\label{eqn:veiling}
F_{\textrm{\small{corr}}} = F - 0.22\times F_{\textrm{\small{6100\AA}}} \times B(\lambda,T)/B(\textrm{6100 \AA},T)
\end{equation}
where $B(\lambda,T)$ is the Planck blackbody function and $T=2800$~K. With the correction applied the continuum and Li~\textsc{i}~6708~\AA\ EW once again match the quiescent levels and the only excess is now due to line emission from H$\alpha$, He~\textsc{i} and Na~\textsc{d}. The H$\alpha$ EW is increased from $-$20~\AA\ to the $-$27~\AA\ given in Table~\ref{table:obs}. Given the low temperature of the veiling blackbody the origin of the veiling emission is most likely warm circumstellar dust rather than an accretion shock, which generally emits at much higher temperatures (5000--20000~K). 2MASS~0801 appears to have an excess of $\sim$0.15~mag in $J-H$ and $\sim$0.05~mag in $H-K_{s}$ relative to 2MASS~0820 and other mid-M members which could indicate the presence of circumstellar material. The star inhabits a region of colour space close to two strong accretors known to possess disks -- RECX~11 and ECHA~J0843.3 have Class~II SEDs, with significant excess at wavelengths greater than 3~$\mu$m \citep{Sicilia-Aguilar09}. 

Reddening (i.e. interstellar dust) could also explain the colour excesses. The \citet{Schlegel98} reddening along a line of sight to 2MASS~0801 is $E(B-V)=0.4$~mag and the star sits on a ridge of prominent dust emission in the \textit{IRAS} maps. It is possible that the stellar colours are affected by reddening if the dust lies in front of the star. From optical polarisation studies the distance to the ridge has been estimated to be $\sim$115~pc \citep{Cleary79}, 40\% larger than the 83~pc estimated dynamical distance to 2MASS~0801 from Paper~\textsc{i}. An improved distance to the star will be necessary to resolve the reddening issue.

\subsection{Radial velocity variations}
\label{sect:rv}

Our multi-epoch observations provide insight into any radial velocity variations in the two stars. The measurement of velocities from the \textit{WiFeS} spectra is described in Paper~\textsc{i} and the resultant RV values are given in Table~\ref{table:obs}. From 260 observations of M spectral type velocity standards we have established that \textit{WiFeS} is capable of an RMS RV precision of 2~\kms. The standard deviation of the 13 RV measurements of 2MASS~0820 is 1.9 \kms, consistent with a flat velocity curve. The RV from the \textit{MIKE} measurement, 17.4 $\pm$0.5~\kms, is also consistent with the mean \textit{WiFeS} value of 18.4~\kms\ within the errors. In contrast, 2MASS~0801 has a standard deviation of 3.3~\kms\ from 9 measurements and an RMS variation nearly five times larger than 2MASS~0820. 

In Paper~\textsc{i} we suggested that in conjunction with its elevated position in the cluster CMD the RV variation of 2MASS~0801 suggested binarity. Any such companion would need to be significantly cooler and fainter as we do not detect a second set of lines at any epoch. The timescale of the RV variation suggests a period of order days to weeks with a velocity amplitude of less than a few tens of \kms. Without high-resolution imaging we cannot constrain the orbital semi-major axis other than to say any companion is not resolved by contemporary imaging surveys such as DSS2 and 2MASS. High resolution spectroscopy and adaptive optics imaging \citep[e.g.][]{Kohler02} will be necessary to resolve any companion. 

Another possibility is that the RV variation is induced by activity or surface features co-rotating with the star. \citet{Martin06} detect a 3.5~\kms\ amplitude, 3.7~hr periodicity in the optical RV data of the M9 brown dwarf LP~944$-$20, which they attribute to rotationally modulated inhomogeneous surface features. Using a toy model with a single spot 100$-$200~K cooler than the photosphere covering $\sim$10\% of the stellar surface, \citet{Reiners10} can generate a velocity amplitude of a few \kms\ for stars of 2800--5700~K, similar to the signal seen in LP~944$-$20 and 2MASS~0801.  We do not find any correlation between the RV and either the H$\alpha$ EW or $v_{10}$ so any velocity variation is probably driven by rotation not chromospheric flaring. The rotational period of 2MASS~0801 has not been measured but is likely on the order of days, similar to the other late-type members of \echa\ \citep{Lawson01} and the timescale of observed RV variation. 

\section{Discussion}

Our results show that H$\alpha$ variability in $\sim$8~Myr PMS stars can be substantial on both short (hours--days) and long (months) timescales. This variation is probably driven primarily by chromospheric activity, which can generate broad H$\alpha$ profiles mimicking accretion over short timescales. However, we also have evidence for at least one accretion event in 2MASS~0820 which requires confirmatory follow-up observations. Additional mid-IR observations will be necessary to detect the presence of any circumstellar disk around the star feeding the accretion. Assuming the duty-cycle of episodic accretion is low, single-epoch surveys of accreting objects, especially in the critical age range 5--10~Myr when inner disks are being cleared and giant planet formation takes place, are likely missing a large fraction of accreting objects. Gas depletion timescales derived from the fraction of accretors are therefore likely underestimates. This may provide another explanation for the difference in characteristic timescales between mass accretion and dust dissipation found by \citet[][2.3~Myr and 3~My respectively]{Fedele10}, without the need to invoke planet formation and/or migration in the inner disk as a possible mechanism for halting accretion.  A larger survey of the disk and accretion properties of outlying \echa\ members, combined with more detailed investigation of the true accreting fraction of PMS clusters from multi-epoch surveys is needed to resolve this discrepancy.

\section*{Acknowledgments}

Australian access to the Magellan Telescopes was supported through the National Collaborative Research Infrastructure Strategy of the Australian Federal Government. DDRB acknowledges financial support from the Access to Major Research Facilities Program, which is supported by the Commonwealth of Australia under the \textit{International Science Linkages Program}.

\end{document}